# A Structured Construction of Optimal Measurement Matrix for Noiseless Compressed Sensing via Analog Polarization

Linbo Li, Hessam Mahdavifar, and Inyup Kang

*Abstract*—In this paper, we propose a method of structured construction of the optimal measurement matrix for noiseless compressed sensing (CS), which achieves the minimum number of measurements which only needs to be as large as the sparsity of the signal itself to be recovered to guarantee almost error-free recovery, for sufficiently large dimension. To arrive at the results, we employ a duality between noiseless CS and analog coding across sparse additive noisy channel (SANC). Extending Rényi's Information Dimension to Mutual Information Dimension (MID), we show the operational meaning of MID to be the fundamental limit of asymptotically error-free analog transmission across SANC under linear analog encoding constraint. We prove that MID polarizes after analog polar transformation and obeys the same recursive relationship as BEC. We further prove that analog polar encoding can achieve the fundamental limit of achievable dimension rate with vanishing $P_e$ across SANC. From the duality, a structured construction scheme is proposed for the linear measurement matrix which achieves the minimum measurement requirement for noiseless CS.

## I. Introduction

The era of "big data" is enabled by the rapid growth of data acquisition from various sources, such as different kinds of sensor networks, data sampling and collecting devices. A problem of central interest is to reduce the cost of data sampling and improve the accuracy and efficiency of data recovery. The question of accurate data recovery has been answered by the Shannon-Nyquist sampling theorem [1], which states that a signal $f(t)$ can be completely recovered by its uniformly sampled values spaced $\frac{1}{2W}$ apart, if the bandwidth of $f(t)$ is confined to within $W$. However, for many practical situations, the Nyquist sampling rate is too high considering the cost of sampling devices/sensors, or the hardware limitation of sampling rate.

To overcome this problem, compressed sensing (CS) has attracted much research interest with the prospect of drastically reducing the sampling rate while maintaining high signal recovery accuracy [3]—[8]. In CS, a sparse signal can be sampled and reliably recovered with a sampling rate proportional to the underlying content or "sparsity" of the signal, rather than the signal's bandwidth. The sparsity $L$ of an $M$-dimensional vector $\mathbf{e}$ is defined by the number of non-zero elements, or in other words the cardinality of the support of $\mathbf{e}$ as $L = \|\mathbf{e}\|_0 = |\text{supp}(\mathbf{e})|$, where $\text{supp}(\mathbf{e}) = \{i: e_i \neq 0\}$. Vector $\mathbf{e}$ is sparse if $L \ll M$. In CS, the sparse signal $\mathbf{e}$ is sampled by taking $P$ linear measurements of $\mathbf{e}$. Let $\mathbf{F}$ be the $P \times M$ measurement matrix, the measurement output $\mathbf{y} \in \mathbb{R}^P$ is

$$\mathbf{y} = \mathbf{Fe}. \tag{1}$$

The main question to be answered by CS is: What is the minimum number of measurements necessary for reliable recovery of the original sparse signal **x**, or what is the minimum number of linear equations in order to solve the vastly underdetermined linear system as specified by (1)?

It is shown in [9, 10] that the signal **x** is the unique solution to the following $\mathcal{L}_0$-minimization problem

$$\min \|\mathbf{x}\|_0 \quad \text{s.t.} \quad \mathbf{y} = \mathbf{Fx}, \tag{2}$$

provided that the sparsity of **x** satisfies

$$L \leq \frac{\text{spark}(\mathbf{F}) - 1}{2}. \tag{3}$$

spark(**F**) is the *spark* of the measurement matrix **F**, which is the minimum number of its columns which are linearly dependent [16, 17]. As a result, for full-rank **F** the maximum recoverable signal sparsity is

$$L \leq \frac{P}{2}, \text{ or } P \geq 2L. \tag{4}$$

In other words, $P = 2L$ measurements are necessary to perfectly recover signal **x** with sparsity of $L$.

To reduce the prohibitive complexity of $\mathcal{L}_0$-minimization (NP-hard as in [11]), $\mathcal{L}_1$ convex relaxation is used, which goes by the name of basis pursuit [18]. According to [9, Theorem 1.3], if certain RIP (restricted isometry property) condition is satisfied, noiseless CS can be exactly solved by the $\mathcal{L}_1$ convex optimization (which can be recast as a linear program), and it admits the same unique solution as the original $\mathcal{L}_0$-minimization. We note that the result in [9] is deterministic and guarantees exact recovery for *all* sparse signals *non-asymptotically* under suitable conditions. In contrast, our paper takes an information-theoretic approach and our results are derived in the large block-length regime. Our main interest is to find the information-theoretic limit or "capacity" of sparse signal recovery. The aim is not to perfectly recover all the sparse signals. Rather, our aim is to recover *almost all* the sparse signals with vanishing error probability for large block-length.

In [14, 15] it is shown that the equivalence between $\mathcal{L}_0$ and $\mathcal{L}_1$ holds true with overwhelming probability for certain classes of random matrices. In addition, [9] showed that the Gaussian unitary ensemble, which is a class of random matrices composed of i.i.d. Gaussian entries, satisfies the RIP condition with overwhelming probability. The matrix multiplication by **F** in (1) can be thought of as a form of "encoding" or "compressing" the source vector **x**. Analogous to classical channel coding, random encoding matrix suffers from the problem of high encoding complexity, whereas a *structured* encoding matrix is of great practical interest since it lends itself to lower implementation complexity.

In this paper, we propose a structured construction of the optimal measurement matrix for noiseless CS, under which we prove that asymptotically error-free recovery is attained for large block size. We will show that the number of linear measurements of our structured construction only needs to be as large as the sparsity of the signal itself to guarantee asymptotically error-free recovery. This result reduces the number of measurements in (4) by half, and achieves the minimum measurements requirement for given signal sparsity in [2].

Our line of reasoning and paper organization go as follows. In Section II, a duality is utilized to transform measurement requirement of noiseless CS to the transmission rate of analog sparse noisy channel. Then in Section III, in order to solve the fundamental limit of analog sparse noisy transmission, following Rényi's original definition of information dimension for real-valued random variables and vectors [13], we extend Rényi information dimension to the cases of conditional entropy and mutual information, and term these quantities *Conditional Information Dimension* (CID) and *Mutual Information Dimension* (MID), respectively. We note that [2] showed Rényi information dimension as the fundamental limit of almost lossless analog source compression. The authors of [19] further showed that, with linearity of analog compressor, the optimal phase-transition threshold of measurement rate for reconstruction error probability can be characterized by Rényi information dimension and MMSE dimension. In particular, for discrete-continuous mixture input which is most relevant in CS practice, [19] showed the optimal phase-transition for measurement rate is given by the weight of the continuous part, regardless of the distribution of the discrete and continuous components. In this paper, using the duality in Section II we focus our attention on SANC (sparse additive noisy channel), which will be analyzed in our analog polar construction. We will show that the operational meaning of MID is the fundamental limit of asymptotically error-free analog transmission under certain conditions on the channel transition distribution $P_{y|x}$ and linear encoder constraint. The performance limit is in terms of the asymptotic ratio of the dimensionality of the codebook ($N$) to the codeword length ($M$) as $M \to \infty$, where the codebook is an $N$-dimensional subspace of $\mathbb{R}^M$. This ratio is termed *Achievable Dimension Rate* in this paper. Following the performance limit of analog sparse noisy channel, the minimum number of measurements to guarantee asymptotically error-free sparse recovery in noiseless CS can be equivalently obtained using the duality. In Section IV we will discuss the channel polarization effect for the sparse noisy channel of interest. We prove that analog polar encoding can achieve the fundamental limit of achievable dimension rate with vanishing $P_e$ across SANC. In Section V we provide an explicit structured construction of the linear measurement matrix which achieves the fundamental performance limit, using machinery of channel polarization. Numerical experiments are provided in Section VI to demonstrate the performance of analog polar encoding under a practical decoder and compare it with the capacity-achieving polar decoder in the proof of Theorem 2, for finite dimension. Section VII contains our concluding remarks.

Existing works in the literature that are related to our results are discussed here. Inspired by the application of second order Reed-Muller code in CS, [20] proposes to use polar code as deterministic measurement matrix in CS, and compares the numerical performances of Polar and Gaussian measurement matrices using practical BPDN (Basis Pursuit DeNoising) algorithm. However, it is not discussed in [20] the fundamental limit of CS and the optimality of polar construction in terms of information dimension limit. In [22], the results on universal source polar coding are applied to compressively sensing and recovering a sparse signal valued on a finite field. [21] proposes CS sensing scheme using deterministic partial Hadamard matrices (selecting a subset of rows of the original Hadamard matrix) and proves that it is lossless and entropy preserving for discretely distributed random vectors with a given distribution. Expander graphs are used in construction of random sensing matrices with RIP and requiring reduced degree of randomness in [23]. In [24], different deterministic sensing matrices (including ones formed from discrete chirps, Delsarte-Goethals codes, extended BCH codes) are shown to achieve Uniqueness-guaranteed Statistical RIP (UStRIP). The spatial coupling idea in coding theory is explored in [25] and [26] to utilize spatial coupling matrix for sensing and it is shown to achieve information dimension limit. However, the sensing matrix used in [25] and [26] is still randomly generated among an ensemble. We also note the independent work in [27] showing the polarization of Rényi information dimension (RID) of i.i.d. sequence of mixture random variables which is transformed by Hadamard matrix. However, a decoding scheme is not proposed in [27] and it did not show that asymptotically error-free reconstruction can be achieved from the RID polarizing Hadamard transformation.

We note that our derivation is information-theoretic in nature. Thus all the results hold in the asymptotic regime of large block length. For the sake of simplicity, we consider real-valued signal and noise in this paper. It is straightforward to generalize all results to the field of complex numbers.

## II. Duality between Noiseless Compressed Sensing and Analog Transmission across Sparse Noisy Channel

As in [9], let us consider the following model of analog encoding and transmission across noisy channel

$$\mathbf{y} = \mathbf{A}\mathbf{x} + \mathbf{e}, \tag{5}$$

where $\mathbf{x} \in \mathbb{R}^N$ is the signal vector, $\mathbf{A}$ is an $M \times N$ ($M > N$) full-rank analog code generator matrix, $\mathbf{e} \in \mathbb{R}^M$ is the noise vector, and $\mathbf{y} \in \mathbb{R}^M$ is the channel output.

We will utilize a duality between noiseless CS (1) and analog coding and transmission (5), so that the minimum measurement for CS is equivalently transformed to a problem of rate limit of analog transmission subject to linear encoding constraint. The precise definition of analog transmission rate is deferred to Definition 5 in Section III. The minimum measurement requirement for CS is obtained as a result of the duality.

Denote by $\mathbf{F}$ the projection matrix onto the kernel of matrix $\mathbf{A}$ (the annihilator of $\mathbf{A}$ as considered in [9]). The dimensionality of $\mathbf{F}$ is $(M-N) \times M$, and it satisfies $\mathbf{F}\mathbf{A} = \mathbf{0}$. Therefore

$$\mathbf{y}' \triangleq \mathbf{F}\mathbf{y} = \mathbf{F}\mathbf{e}. \tag{6}$$

The ratio of the lengths of information signal and transmitted signal is $\frac{N}{M}$ in (5). For sparse noise vector $\mathbf{e}$, it can be seen that analog coding subject to sparse additive noisy channel (SANC) in (5) can be converted to noiseless CS in (6). From (5), it is evident that if $\mathbf{e}$ can be exactly recovered by CS, the information signal $\mathbf{x}$ can be recovered error-free by subtracting $\mathbf{e}$ from (5).

Conversely, for a noiseless CS problem

$$\mathbf{y}' = \mathbf{Fe},$$

where $\mathbf{e} \in \mathbb{R}^M$ is a sparse vector and $\mathbf{F}$ is the $(M-N) \times M$ measurement matrix. First, we find any vector $\mathbf{y}$ that satisfies $\mathbf{Fy} = \mathbf{y}'$. Let us note that $\mathbf{y}$ is not constrained to be sparse. As a result,

$$\mathbf{F}(\mathbf{y} - \mathbf{e}) = \mathbf{0}. \tag{7}$$

From (7) $(\mathbf{y} - \mathbf{e})$ must belong to the subspace spanned by the columns of $\mathbf{A}$, where $\mathbf{A}$ is the matrix that spans the null space of matrix $\mathbf{F}$. Therefore, there exist $\mathbf{x}$ such that $(\mathbf{y} - \mathbf{e})$ can be represented as $\mathbf{y} - \mathbf{e} = \mathbf{Ax}$. As a result,

$$\mathbf{y} = \mathbf{Ax} + \mathbf{e}. \tag{8}$$

Therefore, noiseless CS is transformed to linear analog coding across sparse noisy channel $\mathbf{e}$.

After decoding of the information signal $\mathbf{x}$, the sparse vector $\mathbf{e}$ can be perfectly recovered by $\mathbf{e} = \mathbf{y} - \mathbf{Ax}$. As a result of the duality, maximizing the analog transmission rate (precise definition deferred to Section III) subject to linear encoding constraint will minimize the measurement requirement for the corresponding noiseless CS problem. In Section III we will calculate the information-theoretic limit of linear analog encoding across SANC, and design the corresponding linear encoding scheme in subsequent sections.

*Remark 1*: It is useful to draw the analogy of analog coding of real-valued signal to the classical block coding on the finite field $\mathbb{F}_q$, where q is a power of a prime number. For linear encoding of the vector $\mathbf{x}$ by a generator matrix $\mathbf{A}$ followed by transmission over a channel represented by the additive error vector $\mathbf{e}$, the received vector is $\mathbf{y} = \mathbf{Ax} + \mathbf{e}$. A linear encoding scheme picked from the family of maximum distance separable (MDS) codes can correct up to $\frac{M-N}{2}$ errors. Reed-Solomon code is such an example. Let $\mathbf{F}$ be the parity check matrix with dimensionality $(M-N) \times M$, which satisfies $\mathbf{FA} = \mathbf{0}$. We have

$$\mathbf{y}' \triangleq \mathbf{Fy} = \mathbf{Fe}.$$

$\mathbf{y}'$ is the decoding syndrome. For $\mathbf{F}$ with full rank, its spark is given by $\text{spark}(\mathbf{F}) = M - N + 1$. Therefore from the discussion in Section I, $\mathbf{e}$ can be perfectly recovered if its support satisfies

$$\|\mathbf{e}\|_0 \leq \frac{\text{spark}(\mathbf{F}) - 1}{2} = \frac{M-N}{2}. \tag{9}$$

In turn, the information vector $\mathbf{x}$ can be reliably decoded. Finally, let us note that it is well-known that the RS code does not achieve the channel capacity. As a result, the upper bound in (9) on $\mathbf{e}$'s sparsity cannot be optimal. Analogous to the $\mathcal{L}_0$ scheme and the upper bound in (3) for CS, on finite field, RS code is a *finite* block-length scheme and guarantees perfect recovery for *all* sparse signals with sparsity up to $\frac{M-N}{2}$. It has been shown that with better construction of the encoding matrix (binary polar code is an example [12]), decoding error can be made arbitrarily small and channel capacity can be achieved with asymptotically large codeword length. In subsequent sections, we will apply this analogy to real field.

III. INFORMATION-THEORETIC LIMIT OF ANALOG TRANSMISSION ACROSS SPARSE NOISY CHANNEL

*A. Definitions and Modeling*

From the duality discussed in Section II, we are interested in SANC as described in (5). We will find the information-theoretic limit of its transmission rate. A structured rate-limit-achieving linear encoding scheme will also be proposed by utilizing the channel polarization effect. First of all, the following modeling and definitions need to be introduced to accurately define our framework. $SANC(p)$ with Gaussian noise is modelled as

$$y = x + \alpha n, \tag{10}$$

where $n$ is a Gaussian noise component with 0 mean and variance $\sigma^2$, and $\alpha$ has the following distribution:

$$\alpha = \begin{cases} 0, & \text{w.p. } (1-p) \\ 1, & \text{w.p. } p \end{cases}. \tag{11}$$

$\alpha$ is independent of the noise $n$. For the channel in (10), it is easy to verify that the conditional distribution of $y$ given $x$ is given by

$$P_{y|x}(y|x) = (1-p)\delta(y-x) + p \cdot \mathcal{N}(x, \sigma^2), \tag{12}$$

where $\mathcal{N}(x, \sigma^2)$ denotes the Gaussian distribution of mean $x$ and variance $\sigma^2$. The question we need to answer is: What is the maximum rate of reliably transmitting real-valued signals across this channel in (10)? In order to answer the question above, we will need to accurately define the quantities of interest. To this end, some new definitions will be introduced first.

Let us extend Rényi's definition of information dimension to the cases of conditional entropy and mutual information. We term these quantities CID and MID, respectively. Let us recall the definition of the original Rényi information dimension as:

*Definition 1* (Rényi Information Dimension [13]): Let $X$ be a real-valued random variable. For integer $m$, define the following quantized random variable

$$X_{(m)} = \frac{\lfloor mX \rfloor}{m}.$$

The Rényi information dimension is defined as

$$d(X) = \lim_{m \to \infty} \frac{H(X_{(m)})}{\log m},$$

if the limit above exists. Alternatively, Rényi information dimension has the following equivalent definition.

*Definition 2* (Equivalent Definition of Rényi information dimension [2]): Let us consider the following mesh cube of size $\varepsilon$ on $\mathbb{R}^k$:

$$C_{\mathbf{z},\varepsilon} = \prod_{j=1}^{k} [z_j \varepsilon, (z_j + 1)\varepsilon),$$

where $\mathbf{z} = (z_1, \dots z_k)$ is the k-dimensional integer vector of cube index. The entire $\mathbb{R}^k$ space is divided into mesh cubes of size $\varepsilon$ across all possible values of $\mathbf{z}$. For an arbitrary $k$-dimensional real-valued random vector $\mathbf{X}$, its Rényi information dimension can be equivalently defined as

$$d(\mathbf{X}) = \lim_{\varepsilon \to 0} \frac{H(\mu_{(\varepsilon)})}{\log \frac{1}{\varepsilon}},$$

where $\mu_{(\varepsilon)}$ is the discrete probability measure by setting

$$\mu_{(\varepsilon)}(\mathbf{z}) = \mu(C_{\mathbf{z},\varepsilon}),$$

where $\mu(\cdot)$ is the original probability measure of random vector $\mathbf{X}$.

Now we extend the Rényi information dimension to the cases of condition entropy and mutual information. Here we adopt the $\varepsilon$-quantization form in our definitions.

*Definition 3* (*Conditional Information Dimension*): For random variables $X$ and $Y$, we define the following CID if the respective limit exists:

$$d(Y|X) = \lim_{\varepsilon \to 0} \frac{H(Y_{(\varepsilon)}|X_{(\varepsilon)})}{\log \frac{1}{\varepsilon}} = \lim_{\varepsilon \to 0} \frac{H(Y_{(\varepsilon)}, X_{(\varepsilon)})}{\log \frac{1}{\varepsilon}} - \lim_{\varepsilon \to 0} \frac{H(X_{(\varepsilon)})}{\log \frac{1}{\varepsilon}},$$

where $Y_{(\varepsilon)}$ and $X_{(\varepsilon)}$ are the quantized random variables of $Y$ and $X$ by the mesh cubes of size $\varepsilon$.

*Definition 4* (*Mutual Information Dimension*): For random variables $X$ and $Y$, the MID of $X$ and $Y$ is defined as the following, if the respective limits $d(Y)$ and $d(Y|X)$ exist

$$d(Y;X) = d(Y) - d(Y|X).$$

It is easy to verify that if the limits $d(X)$ and $d(X|Y)$ exist, $d(Y;X) = d(X;Y) = d(X) - d(X|Y)$. $d(Y;X)$ across SANC is given by the following result.

*Proposition 1*. With absolutely continuous input, the MID across $SANC(p)$ is given by

$$d(Y;X) = 1 - p. \tag{13}$$

Proof. In Appendix I.

### B. Operational Meaning of Mutual Information Dimension

In this section, we will find the following operational meaning of MID as defined in Definition 4:

- MID is the fundamental limit of *asymptotically error-free analog transmission* (rigorous definition provided in Definition 5) under certain conditions on the channel transition distribution $P_{y|x}$ and under the regularity constraint of linear analog encoder. The performance limit is in terms of the asymptotic ratio of the dimensionality of the codebook ($N$) to the codeword length ($M$) as $M \to \infty$. The codebook is an $N$-dimensional subspace of $\mathbb{R}^M$, using which asymptotically error-free analog transmission is achieved. The ratio is termed *Achievable Dimension Rate* in this paper.

In what follows, we will make the statement above accurate. We impose linear encoding constraint on the transmitter for two reasons. First, with the codebook taking a continuum of entries, it would be impossible to discuss meaningfully the fundamental limit of real-valued data transmission without imposing certain regularity constraint on the encoder. For example, it is well-known that there is a one-to-one mapping (although a highly discontinuous and irregular one) from $\mathbb{R}$ to $\mathbb{R}^n$. Therefore, without any constraint on the encoding scheme, a single real number could be used to encode a whole vector of real numbers, which would result in infinite transmission capacity.

Second, linear encoding scheme appears naturally from our discussion in Section II on the duality between noiseless CS and analog data transmission.

*Definition 5* (*Achievable Dimension Rate* of Asymptotically Error-Free Analog Transmission): Let $P_{y|x}(y|x)$ be the channel transition distribution of a real-valued analog transmission channel.

- Codebook: For codeword of length $M$, given linear encoding constraint, the codebook is an $N$-dimensional ($N \leq M$) subspace of $\mathbb{R}^M$, which is denoted by $\mathbf{V} \subset \mathbb{R}^M$.
- Data transmission: For codeword $\mathbf{x_0} \in \mathbf{V}$, the $M$-dimensional channel output $\mathbf{y}$ is generated according to repeated and independent applications of $P_{y|x}(y|x)$ componentwise to the entries in $\mathbf{x_0}$.
- Decoder: The decoder is a function $g: R^M \to R^M$ that generates $\hat{\mathbf{x}} = g(\mathbf{y})$.
- Decoder error probability: $P_e^{(M)} = \Pr(g(\mathbf{y}) \neq \mathbf{x_0})$.
- *Achievable Dimension Rate*: A dimension rate $R$ is said to be *achievable* if there exists a sequence of $N = \lceil RM \rceil$-dimensional codes such that the error probability $P_e^{(M)} \to 0$ as the codeword length $M \to \infty$.

Regarding the achievable dimension rate, the following result holds for i.i.d. uses of $SANC(p)$, namely, the sparse noise $\mathbf{e}$ is i.i.d. according to the probabilistic model in (11):

*Theorem 1*. Across $SANC(p)$ and subject to linear encoding constraint, a dimension rate $R$ is achievable, as $M \to \infty$ with $P_e^{(M)} \to 0$, if

$$R < (1-p).$$

Conversely, subject to linear encoding constraint, any sequence of $N = \lceil RM \rceil$-dimensional codes with $P_e^{(M)} \to 0$ must have the dimension rate $R$ satisfying

$$R \leq 1 - p.$$

Proof. This can be readily proved by invoking the duality in Section II and [2, Theorem 6] for linear analog compression of discrete-continuous sources.

From the proof of [2, Theorem 6], for sufficiently large $M$, there exists at least one realization of codebook $\mathbf{V}^{(M)}$ such that the dimension rate $R < (1-p)$ is achievable with arbitrarily small decoding error probability. The existence of a sequence of good codebooks $\mathbf{V}^{(M)}$ for analog transmission prompts us to find a structured construction of it for practical purposes. Based on Theorem 1 and the duality between noiseless CS and analog transmission, for sufficiently large $M$, there exists a sequence of measurement matrix $\mathbf{F}^{(M)}$ such that the number of measurement $\lceil \rho M \rceil$ is achievable with arbitrarily small recovery error probability for $\rho > p$.

*Remark 2*: We provide an interesting interpretation of (13). It can also be obtained by means of adding a vanishing Gaussian noise term $n_e$ to (10) and the noise free channel $y = x$, and evaluate their respective rate of growth of the mutual information as the variance of $n_e$ goes to zero.

Specifically, we add a vanishing Gaussian noise term $n_e$, which is independent of $x, \alpha,$ and $n$ and has variance $\sigma_e^2$, to both (10) and the noise free channel:

$$y = x + \alpha n + n_e \qquad (14)$$
$$y = x + n_e \qquad (15)$$

In the regime of vanishing $\sigma_e^2$, the mutual information of channel (15) behaves as

$$\frac{I_{\sigma_e^2}(x;y)}{\log \frac{1}{\sigma_e}} = 1 + o(1), \quad \text{as } \sigma_e^2 \to 0. \qquad (16)$$

For the channel in (14), the mutual information is

$$I(x;y) = h(y) - h(y|x).$$

$h(\cdot)$ denotes the differential entropy. Since $h(y)$ is finite, we only need to calculate the second term $-h(y|x)$. Given the conditional probability distribution

$$P_{y|x}(y|x) = (1-p)\mathcal{N}(x, \sigma_e^2) + p\mathcal{N}(x, \sigma_e^2 + \sigma^2), \qquad (17)$$

it is readily shown that the mutual information behaves as

$$\frac{I_{\sigma_e^2}(x;y)}{\log\frac{1}{\sigma_e}} = \frac{-h(y|x)}{\log\frac{1}{\sigma_e}} = (1-p) + o(1), \quad \text{as } \sigma_e^2 \to 0. \qquad (18)$$

As a result, for the two channels in (14) and (15), the relative ratio of the rate of growth of the mutual information is $(1-p)$ as $\sigma_e^2 \to 0$, which turns out to be the same as the MID in (13). Therefore, in this case in the asymptotic domain of vanishing noise, the capacity of transmitting discrete bits is related to the capacity of transmitting real-valued signals through its asymptotic rate of growth.

## IV. POLARIZATION OF SPARSE NOISY CHANNEL

### A. SANC and SAEC

From our discussion in Section III, there exists a sequence of codebook **V** which asymptotically achieves the dimension rate of $(1-p)$ for the channel model in (10). Now a natural question to ask is: What is a structured way of linear encoding in order to achieve the dimension rate in Theorem 1? In this section, we will answer this question with the tool of channel polarization.

First, let us note that according to (13), the MID of the channel in (10) does not depend on the distribution of the noise component $n$, namely, its variance for the Gaussian case we have assumed. It depends only on the probability of noisy realization of the channel itself (the $p$ parameter). We recognize this as an erasing effect, namely, the input real-valued signal is erased by a noisy realization of the channel and cannot possibly be recovered without distortion. Therefore, what really matters is $p$, which indicates the probability of noisy channel realization, instead of the distribution of the noise component. Based on this observation, the channel in (10) is equivalent to the following sparse analog erasure channel (SAEC) in terms of MID.

For SAEC, with input $x$, the output $y$ is given by

$$y = \begin{cases} n, & \text{w. p. } p, \text{where } n \text{ is WGN component} \\ x, & \text{w. p. } (1-p) \end{cases}. \qquad (19)$$

Here we can regard the parameter $p$ as the erasure probability (also the noise sparsity), as analogous to the erasure probability of the binary erasure channel (BEC).

*Remark 3*: We will use SAEC channel model in (19) in MID derivation, since it lends itself to simpler analysis. However, we note that the same MID results will hold for SANC in (10) as long as they have the same noise sparsity $p$.

### B. Channel Polarization and Duality between SAEC and BEC

In this section, we derive a duality between the SAEC and BEC in terms of channel polarization with respect to the MID and Shannon capacity, respectively. Reference [12] proved that for BEC, through recursive application of the following one-step transformation on the channel input

$$\begin{pmatrix} 1 & 1 \\ 0 & 1 \end{pmatrix}, \qquad (20)$$

channel polarization is achieved in the limit of large input size. Also, the one-step transformation preserves the combined channel capacity of the two new bit channels. If we denote the erasure probability by $\varepsilon$ for the original BEC, the erasure probability for the new channel pair after the one-step transformation is given by [12]:

$$2\varepsilon - \varepsilon^2 \text{ and } \varepsilon^2.$$

Therefore, after one-step transformation, one of the new bit channel becomes better and the other one gets worse.

For SAEC on real-valued signals, we will show that the following one-step transformation in (21) on independent input $x_1$ and $x_2$ will produce the same polarization effect in terms of MID:

$$\begin{pmatrix} z_1 \\ z_2 \end{pmatrix} = \begin{pmatrix} \beta & \beta \\ 0 & 1 \end{pmatrix} \begin{pmatrix} x_1 \\ x_2 \end{pmatrix}. \qquad (21)$$

We can choose, for example, $\beta = 1$. After the transformation in (21), $z_1$ and $z_2$ are sent through two independent realizations of SAEC and the corresponding output is denoted by $y_1$ and $y_2$, as illustrated in Fig. 1.

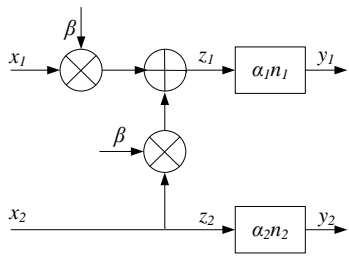

Fig. 1. One-step transformation.

The above linear transformation creates a pair of new channels for the original input $x_1$ and $x_2$. For SAEC, Table I lists the possible channel realizations with the corresponding probability.

TABLE I
CHANNEL REALIZATIONS AND THE CORRESPONDING PROBABILITY

| SAEC REALIZATION | PROBABILITY |
|---|---|
| $y_1 = \beta(x_1 + x_2)$ $y_2 = x_2$ | $(1-p)^2$ |
| $y_1 = n_1$ $y_2 = x_2$ | $p(1-p)$ |
| $y_1 = \beta(x_1 + x_2)$ $y_2 = n_2$ | $p(1-p)$ |
| $y_1 = n_1$ $y_2 = n_2$ | $p^2$ |

In what follows, we will evaluate the MID of the new channel pair through $d(Y_1, Y_2; X_1)$ and $d(Y_1, Y_2, X_1; X_2)$.

*Proposition 2*. After the single-step transformation of (21)

$$MID_1 = d(Y_1, Y_2; X_1) = (1-p)^2, \qquad (22)$$
$$MID_2 = d(Y_1, Y_2, X_1; X_2) = 1 - p^2. \qquad (23)$$

Proof. In Appendix II.

From (22) and (23), it is easy to see that

$$MID_2 > MID_1, \quad if\ 0 < p < 1.$$

Also we notice that

$$MID_1 + MID_2 = 2(1-p),$$

which is exactly the combined MID of the two original SAEC. Thus, the one-step linear transformation of

$$\mathbf{G}_0 = \begin{pmatrix} \beta & \beta \\ 0 & 1 \end{pmatrix} \qquad (24)$$

preserves the combined MID while improving the new channel for $x_2$ and degrading the new channel for $x_1$.

Table II compares our results for SAEC on real field and the results in [12] for BEC on binary field:

TABLE II
COMPARISON OF SAEC AND BEC BEFORE AND AFTER ONE-STEP TRANSFORMATION

| SAEC | BEC |
|---|---|
| Noise sparsity: $p$ | Erasure probability: $\varepsilon$ |
| MID: $1-p$ | Shannon capacity: $1-\varepsilon$ |
| One-step transformation preserves combined MID | One-step transformation preserves combined Shannon capacity |
| After one-step transformation, the new channel pair's MID is: $(1-p)^2$ $1-p^2$ | After one-step transformation, the new channel pair's Shannon capacity is: $(1-\varepsilon)^2$ $1-\varepsilon^2$ |

From Table II, we can see that the properties of the one-step transformation (21) of SAEC are equivalent to those of the one-step transformation (20) of BEC. The noise sparsity $p$ of SAEC serves as the counterpart of erasure probability $\varepsilon$ of BEC.

Then as in [12], we recursively apply the one-step transformation. For the $(n+1)$-th step, the channel construction is illustrated in Fig. 2.

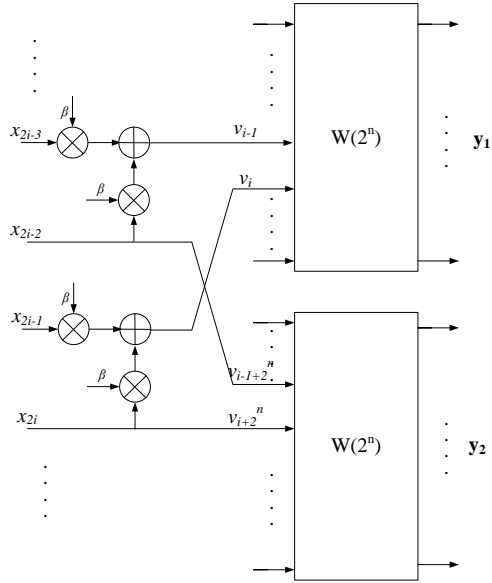

Fig. 2. $(n+1)$-th step of iterative application of the one-step transformation.

In Fig. 2, $W(2^n)$ denotes the channel block formed in the $n$-th step. Let us define

$$MID_{2^n}^{(i)} = d(\mathbf{y}, x_1^{i-1}; x_i),$$

where $x_1^{i-1} = (x_1, \ldots, x_{i-1})$ and $\mathbf{y} = (\mathbf{y}_1^T, \mathbf{y}_2^T)^T$. Following the induction in [12], the same iterative relationship as in [12] for BEC holds:

*Proposition 3.*

$$MID_{2^{n+1}}^{(2i-1)} = \left(MID_{2^n}^{(i)}\right)^2,$$
$$MID_{2^{n+1}}^{(2i)} = 2MID_{2^n}^{(i)} - \left(MID_{2^n}^{(i)}\right)^2.$$

Proof. In Appendix III.

Therefore, following the argument in [12], channel polarization effect will take place for large $n$ in the sense that: The new effective channels will polarize into a class of "good" channels with MIDs approaching 1, and a class of "bad" channels with MIDs approaching 0. The proportion of the good channels is equal to $(1-p)$.

Fig. 3 and Fig. 4 show the channel polarizing effect for block length of $M = 2^{14}$ and sparsity $p = 0.1$. Fig. 3 shows MID of the sub-channels with natural ordering of the sub-channels, whereas Fig. 4 shows the sorted MID in decreasing order. As $M \to \infty$, all MIDs will converge to either 0 (bad channels) or 1 (good channels), except for a subset with vanishing proportion.

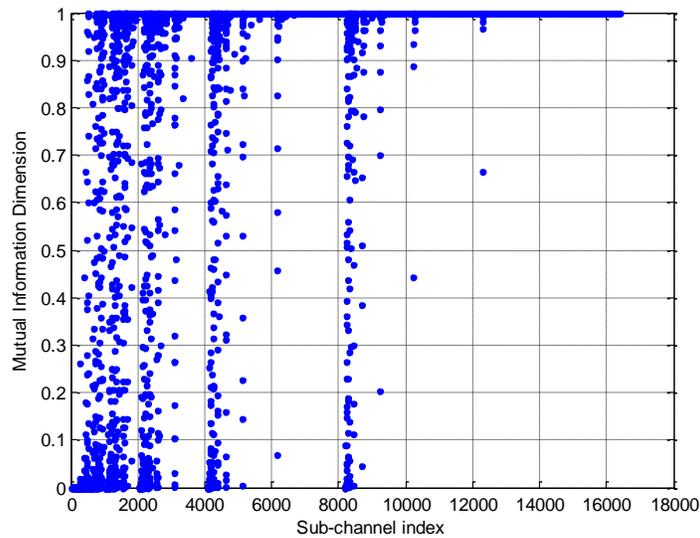

Fig. 3. MID for $p = 0.1$ and $M = 16384$.

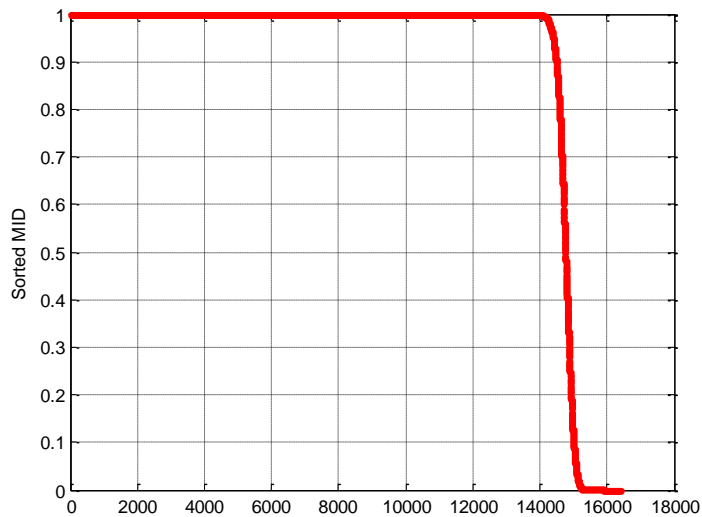

Fig. 4. Sorted MID for $p = 0.1$ and $M = 16384$.

In the $n$-th step of recursive application of the one-step transformation $\mathbf{G}_0$ in (24), a size $2^n \times 2^n$ transformation matrix is formed. The overall transformation from the original signal $\mathbf{x}$ to the input vector $\mathbf{z}$ of the underlying independent SAECs is:

$$\mathbf{z} = (\mathbf{G}_0^{\otimes n} \cdot \mathbf{B}_{2^n})\mathbf{x}, \qquad (25)$$

where $\mathbf{G}_0^{\otimes n}$ is the $n$-fold Kronecker product of $\mathbf{G}_0$, $\mathbf{B}_{2^n}$ is the $2^n \times 2^n$ bit reversal matrix which permutes the columns of $\mathbf{G}_0^{\otimes n}$ in the bit-reverse order.

From MID polarization, we only need to transmit the real-valued signals on the good channels, transmit independent random signals on the bad channels and reveal it to the receiver, with the hope that the dimension rate $1 - p$ will be achieved by such analog polar encoding. This intuitive procedure will be made precise in the following section.

*Remark 4*: As discussed in Section IV.A, we use SAEC in place of SANC for MID calculation, since they are equivalent in that regard and the analysis is simplified. It can be readily verified that for SANC, for which Table I is replaced by Table III below, Proposition 2 will also hold. Then Proposition 3 follows from the same induction in Appendix III. The derivation will be omitted to save space.

TABLE III
CHANNEL REALIZATIONS AND THE CORRESPONDING PROBABILITY

| SANC REALIZATION | PROBABILITY |
|---|---|
| $y_1 = \beta(x_1 + x_2)$<br>$y_2 = x_2$ | $(1-p)^2$ |
| $y_1 = \beta(x_1 + x_2) + n_1$<br>$y_2 = x_2$ | $p(1-p)$ |
| $y_1 = \beta(x_1 + x_2)$<br>$y_2 = x_2 + n_2$ | $p(1-p)$ |
| $y_1 = \beta(x_1 + x_2) + n_1$<br>$y_2 = x_2 + n_2$ | $p^2$ |

### C. Analog Polar Encoding Achieves Vanishing $P_e$ while Achieving the Dimension Rate $(1-p)$ across $SANC(p)$

In this section, we prove that analog polar encoding can asymptotically achieve the dimension rate $(1 - p)$ with error probability $O(2^{-M^\beta})$ for any $0 < \beta < 0.5$. The polar linear encoding matrix **A** is constructed by selecting the columns of the good channels according to MID within the full $M \times M$ polarization matrix denoted by **H**. According to Proposition 3, $MID$ of $SANC(p)$ obeys the same recursive formula as channel capacity $I$ of $BEC(p)$. As a result, $SANC(p)$ and $BEC(p)$ have the same good channel indices.

Let $L_I$ denote the number of selected columns of **H** with higher MID than $1 - \frac{2^{-M^\beta}}{M}$, namely

$$L_I = \left|\left\{i \in \{1, \dots, M\}: MID_M^{(i)} \geq 1 - \frac{2^{-M^\beta}}{M}\right\}\right|.$$

Notice that this selection of columns is identical to the binary polar code construction for $BEC(p)$, denoted by $C_{p,\beta}$, under the same criteria on the symmetric capacity $I \geq 1 - \frac{2^{-M^\beta}}{M}$. Then by the channel polarization for BEC, the rate of the corresponding polar code $\frac{L_I}{M}$ approaches $1 - p$ while the decoding error is upper bounded by $2^{-M^\beta}$ [28]. The following theorem is our result on analog polar encoding across $SANC(p)$.

*Theorem 2*. For $SANC(p)$, for any $0 < \beta < 0.5$, the analog polar encoding scheme described above can asymptotically achieve the dimension rate of $1 - p$ with decoding error probability $O(2^{-M^\beta})$ as the codeword length $M \to \infty$.

Proof. In Appendix IV.

Considering Theorem 1, analog polar encoding achieves maximal achievable dimension rate among all linear encoders. At the end of this section, let us compare our result in Theorem 2 with the rate achieved with prior work. From the duality in Section II, let us transform the problem of noisy analog transmission with linear encoding $\mathbf{y} = \mathbf{Ax} + \mathbf{e}$ to a CS problem $\mathbf{y}' \triangleq \mathbf{Fy} = \mathbf{Fe}$. From the prior result on CS (see (4) and [9, 10]), if the support of the error vector **e** satisfies

$$\|\mathbf{e}\|_0 \leq \frac{M-N}{2}, \tag{26}$$

error-free recovery of **x** is guaranteed. The rate of the analog coding is $r = \frac{N}{M}$, and the normalized sparsity of the error vector is

$$p = \frac{M-N}{2M} = \frac{1}{2}(1-r).$$

As a result, for the original noisy channel $\mathbf{y} = \mathbf{Ax} + \mathbf{e}$, the supported rate of reliable analog transmission is

$$r = 1 - 2p. \tag{27}$$

It is interesting to compare (27) with the achieved dimension rate $1 - p$ in Theorem 2. The reduction of achieved rate is mainly because, whereas the rate in (27) is guaranteed for finite block length and all codewords, the rate in Theorem 2 is achieved asymptotically as the block length goes to infinity.

### D. SC-SANC Polar Decoding

In the proof of Theorem 2 we use decoding by repetition counting of the outcome. It considers all the $L_I \times L_I$ non-singular sub-matrices and therefore has high complexity. In this section we discuss another decoder based on SC (successive cancellation). However, we note that although being more intuitive, its optimality under SANC is still an open problem.

As usual, let $M = 2^n$. Let us denote the $M$-dimensional input signal by $\mathbf{x}_0$, the $N$-dimensional subset of $\mathbf{x}_0$ corresponding to the good channels by **x**, the $(M-N)$-dimensional subset of $\mathbf{x}_0$ corresponding to the bad channels by $\mathbf{x}_b$, and the $M \times M$ full polar transformation matrix by **H**. From (25),

$$\mathbf{z} = (\mathbf{G}_0^{\otimes n} \cdot \mathbf{B}_{2^n})\mathbf{x}_0 = \mathbf{H}\mathbf{x}_0 = \mathbf{Ax} + \mathbf{A}_b\mathbf{x}_b, \tag{28}$$

where **z** is the input to the underlying SANC, **A** is the $M \times N$ matrix whose columns are associated with the good channels in **H**, and $\mathbf{A}_b$ is the $M \times (M-N)$ matrix whose columns correspond to the bad channels. The channel model is

$$\mathbf{y} = \mathbf{Ax} + \mathbf{A}_b\mathbf{x}_b + \mathbf{n}, \tag{29}$$

where **n** is the additive sparse noise vector and is given by

$$\mathbf{n} = \begin{pmatrix} \alpha_1 n_1 \\ \alpha_2 n_2 \\ \vdots \\ \alpha_M n_M \end{pmatrix}. \tag{30}$$

$\mathbf{x}_b$ is revealed to the receiver. Without a priori knowledge of **x**, it is natural to consider the ML rule, namely, finding input vector **x** maximizing the conditional distribution $p_{\mathbf{y}|\mathbf{x}}(\mathbf{y}|\mathbf{x})$. Analogous to the binary case in [12], SC (successive cancellation) based SANC polar decoder can be formulated as follows. Let $\mathcal{G}$ denote the index set of good channels, and $\mathcal{G}^c$ denote the index set of bad channels.

Description of SC-SANC polar decoder:

---

$$\mathbf{y}^{(0)} = \mathbf{y}.$$

for $i = 1$ to $M$

    if $i \in \mathcal{G}^c$
        /*$x_i$ is known to receiver*/
        $\hat{x}_i = x_i$
    else
        $\hat{x}_i = \arg\max_{x_i} W_M^{(i)}(\mathbf{y}^{(i-1)}|x_i)$
    end;

    $\mathbf{y}^{(i)} = \mathbf{y}^{(i-1)} - \mathbf{h}_i \hat{x}_i.$

end;

---

In the SC decoder, $\mathbf{h}_i$ is the $i^{th}$ column of $\mathbf{H}$. Conditional distribution $W_M^{(i)}(\mathbf{y}^{(i-1)}|x_i)$ is calculated via the recursive relationship in (31) and (32) in Appendix III. For example, it is readily verified that $W_4^{(4)}(\mathbf{y}^{(3)}|x_4)$ is given by:

$$W_4^{(4)}(\mathbf{y}^{(3)}|x_4) = \prod_{j=1}^{4}[(1-p)\delta(y_j^{(3)} - x_4) + p\mathcal{N}_{y|x}(y_j^{(3)}|x_4)],$$

where $\mathcal{N}_{y|x}$ is Gaussian distribution of mean $x$ and variance $\sigma^2$. In the above maximization, it is understood that first order $\delta$-function, if satisfied, is larger than finite term, and higher order $\delta$-function, if satisfied, is larger than lower order ones. Unfortunately, it is still rather complex to find the maximizing $\hat{x}_i$ for each stage, and an efficient algorithm for this task is unknown. In Section VI, we will do some numerical experiments with the proposed analog polar code using $\mathcal{L}_1$ decoder.

## V. Deterministic Structured Construction of Optimal Measurement Matrix for Noiseless CS

Following the discussion in Section IV.C and using duality, it follows naturally that the corresponding sensing matrix $\mathbf{F}$ for CS achieves minimum measurements with vanishing reconstruction error rate. The optimal structured measurement matrix $\mathbf{F}$ can be constructed deterministically as follows:

*Optimal Deterministic Structured Construction*: $\mathbf{F}$ is the projection matrix onto the kernel of matrix $\mathbf{A}$, where $\mathbf{A}$ is the submatrix of $\mathbf{H}$ formed by picking the columns of $\mathbf{H}$ corresponding to the good channels, and $\mathbf{H}$ is the channel-polarizing linear transformation matrix in (25).

Given the optimal construction of the measurement matrix $\mathbf{F}$, the noiseless CS problem $\mathbf{y}' = \mathbf{F}\mathbf{e}$ can be solved following the steps below:

1. As in Section II, converting $\mathbf{y}' = \mathbf{F}\mathbf{e}$ to the equivalent analog decoding problem $\mathbf{y} = \mathbf{A}\mathbf{x} + \mathbf{e}$.
2. Use polar decoding algorithm to decode $\mathbf{x}$ from $\mathbf{y}$
3. Recover $\mathbf{e}$ by

$$\mathbf{e} = \mathbf{y} - \mathbf{A}\mathbf{x}.$$

Let us calculate the sparsity of $\mathbf{e}$ that can be recovered error-free. Since

$$MID = 1 - p,$$

where $p$ is the sparsity of **e**, from the fact that **A** achieves maximum dimension rate of $1 - p$, we have

$$r = \frac{N}{M} = 1 - p.$$

As a result

$$p = \frac{M - N}{M}.$$

This is actually two times as good as the result achieved by (26), which says

$$p = \frac{M - N}{2M}.$$

Here we would like to point out again that our proposed polar CS algorithm achieves error-free recovery asymptotically as the block-length goes to infinity, whereas (26) achieves error-free recovery for finite block length.

## VI. NUMERICAL EXPERIMENTS

In this section, we perform some numerical experiments to demonstrate the performance of analog polar encoding. Since an efficient algorithm for analog polar decoding is not known, instead, we will use the $\mathcal{L}_1$ decoder in [9] for simulation purposes. $\mathcal{L}_1$ decoder solves the following program where **A** is our polar analog coding matrix.

$$(P_{L1-Dec}) \quad \min_{\mathbf{x}} \|\mathbf{y} - \mathbf{A}\mathbf{x}\|_1.$$

For $\mathcal{L}_1$ decoder we also simulate Gaussian random encoding. Furthermore, we will also compare the performance of the capacity-achieving polar decoder in the proof of Theorem 2 with $\mathcal{L}_1$ decoder.

In the numerical result, codeword length is 256. The analog information vector **x** is generated according to i.i.d. standard Gaussian distribution $\mathcal{N}(0,1)$. The non-zero entries of the sparse noise vector are also generated according to i.i.d $\mathcal{N}(0,1)$, with their locations randomly placed among the entries of the codeword. Due to finite precision of numerical simulation, a codeword is declared in error if the average $\mathcal{L}_1$ error per dimension exceeds $10^{-4}$. In Fig. 5, the analog code rate is 0.25, and the codeword error rate is plotted against noise sparsity. In Fig. 6, the noise sparsity is fixed to 0.2, and the codeword error rate is plotted against analog code rate. From the figures, $\mathcal{L}_1$ decoder performs far away from the capacity $1 - p$ at $10^{-4}$ error rate. For example, in Fig. 6 the achieved dimension rate of $\mathcal{L}_1$+Polar is only 0.304 for $p = 0.2$, while $\mathcal{L}_1$+Gaussian only achieves dimension rate of 0.332.

For comparison purposes, we calculate the upper bound of the error rate of the capacity-achieving polar decoder in the proof of Theorem 2. From the proof, we know that the error rate can be upper bounded by $P_1 + P_2$, where $P_1$ is the error rate of the corresponding binary polar code for $BEC(p)$, and $P_2 = \Pr\{L_c = L_I\}$ (using the notation therein). $P_1$ can be upper bounded by summing up the corresponding Bhattacharyya parameter and $P_2$ is a binomial coefficient being directly calculated. From the calculation, the decoder in Theorem 2 achieves a dimension rate of at least 0.5 for $p = 0.2$ at $10^{-4}$ error rate, which is significantly better than $\mathcal{L}_1$ even for the relatively small block length of 256.

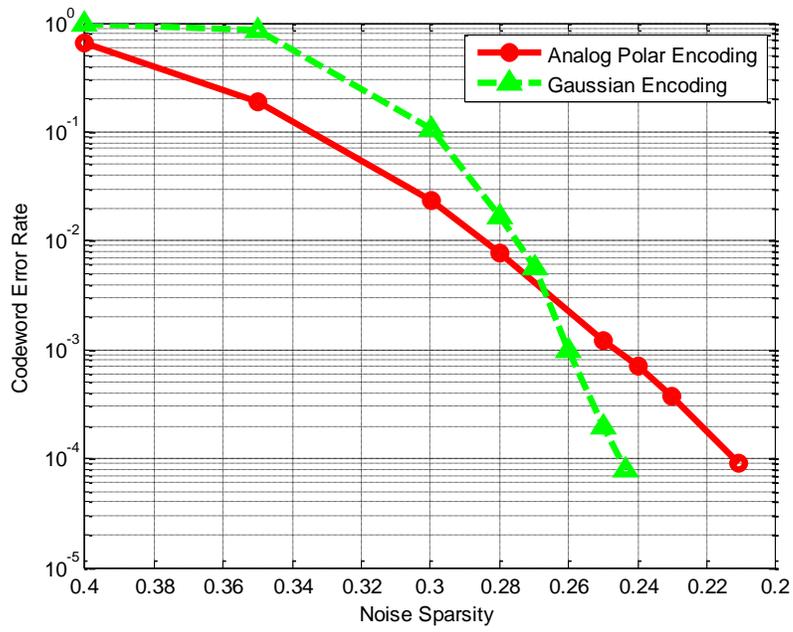

Fig. 5. Codeword ER vs. Noise Sparsity (Analog Code Rate = 0.25).

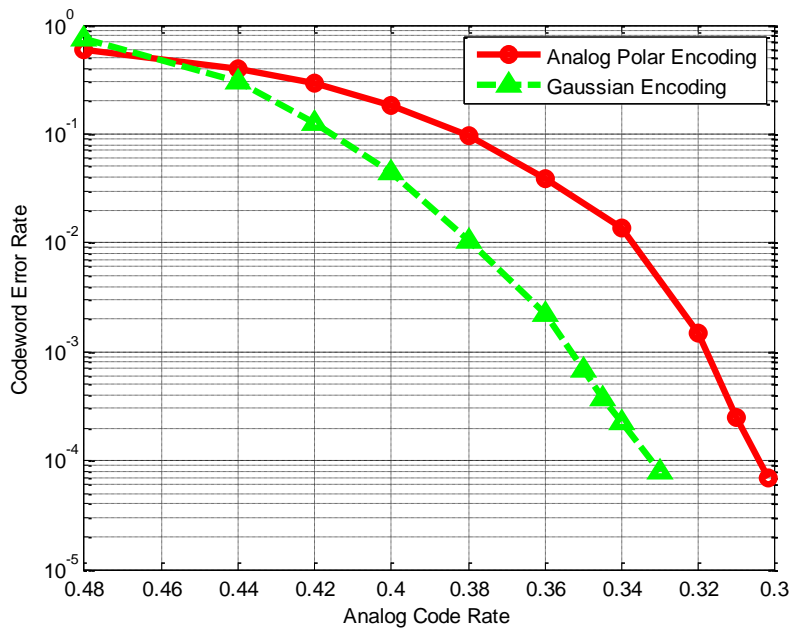

Fig. 6. Codeword ER vs. Analog Code Rate (Noise Sparsity = 0.2).

## VII. CONCLUDING REMARKS

Compressed sensing (CS) and polar coding have been two recent important developments in the areas of signal processing and channel coding. In a conventional sense, there is the fundamental difference between the two fields that CS deals with recovery of real-valued signals, whereas channel coding is concerned with the recovery of discrete bits across noisy channels. However, there also exists profound analogy between the two fields, in that channel coding aims at minimizing the coding redundancy while ensuring reliable communication across noisy channels, whereas in CS we would like to minimize the number of measurements which is required to ensure signal recovery. Hence measurement in CS plays a role that is analogous to the role coding redundancy plays in coding theory.

In this paper, we proposed the concept of mutual information dimension which is important to explore some connection between the two fields. MID has the interpretation of the fundamental limit of achievable dimension rate for analog transmission under certain conditions. We showed that SANC and SAEC lend themselves to the channel polarization effect for large block length, which in turn enables us to propose the deterministic structured construction of the optimal measurement matrix for noiseless CS, which results in the minimum number of linear measurements. We also proved that analog polar encoding can achieve the fundamental limit of achievable dimension rate for SANC with vanishing $P_e$. For our construction, the number of linear measurements only needs to be as large as the sparsity of the signal itself for asymptotically error-free recovery. A duality between noiseless CS and analog coding across sparse noisy channel is utilized in the derivation.

# APPENDIX I

## PROOF OF PROPOSITION 1

The joint distribution $P_{y,x}$ is

$$P_{y,x}(y,x) = P_{y|x}(y|x)P_x(x) = (1-p)\delta(y-x)P_x(x) + p\mathcal{N}(x,\sigma^2)P_x(x).$$

Since $x$ is absolutely continuous, with probability $(1-p)$, $(y,x)$ is absolutely continuous on the 1-dimensional curve $y = x$, and with probability $p$ it is absolutely continuous on $\mathbb{R}^2$. [13] proved the Rényi information dimension of a random variable with discrete-continuous mixed distribution to be the weight of the absolutely continuous part of the distribution, and the results are extended to higher dimensional Euclidean spaces. From [13], the contribution of the first and second terms are $(1-p)$ and $2p$ respectively, and the Rényi information dimension of $P_{y,x}$ is

$$d(Y,X) = (1-p) + 2p = 1 + p.$$

From our definition of CID, we have

$$d(Y|X) = d(Y,X) - d(X) = 1 + p - 1 = p.$$

After integration w.r.t. $x$, it is obvious that the channel output is absolutely continuous. Thus, $d(Y) = 1$. The MID is given by

$$d(Y;X) = d(Y) - d(Y|X) = 1 - p.$$

# APPENDIX II

## PROOF OF PROPOSITION 2

Assuming all inputs and WGNs are absolutely continuous, given

$$\begin{aligned}P_{y_1,y_2|x_1}(y_1,y_2|x_1) &= (1-p)^2 P_{x_2}(y_2)\delta(y_1 - \beta(x_1+y_2)) + (1-p)p \cdot P_{n_1}(y_1)P_{x_2}(y_2) + (1-p)p \\ &\quad \cdot P_{x_2}\left(\frac{y_1}{\beta} - x_1\right)P_{n_2}(y_2) + p^2 \cdot P_{n_1}(y_1)P_{n_2}(y_2)\end{aligned}$$

it is easily shown that the channel output $(y_1, y_2)$ is 2-dimensional absolutely continuous. As a result,

$$d(Y_1, Y_2) = 2,$$

since in this case it coincides with its geometrical dimensionality of 2. The joint distribution of $(y_1, y_2, x_1)$ is

$$\begin{aligned}P_{y_1,y_2,x_1}(y_1,y_2,x_1) &= (1-p)^2 P_{x_2}(y_2)\delta(y_1 - \beta(x_1+y_2))P_{x_1}(x_1) + (1-p)p \cdot P_{n_1}(y_1)P_{x_2}(y_2)P_{x_1}(x_1) + (1-p)p \\ &\quad \cdot P_{x_2}\left(\frac{y_1}{\beta} - x_1\right)P_{n_2}(y_2)P_{x_1}(x_1) + p^2 \cdot P_{n_1}(y_1)P_{n_2}(y_2)P_{x_1}(x_1).\end{aligned}$$

Similar to the derivation in Appendix I, the Rényi information dimension is calculated as:

$$d(Y_1, Y_2, X_1) = 2(1-p)^2 + 3(1-p)p + 3(1-p)p + 3p^2 = 2 + 2p - p^2.$$

So the CID is given by,

$$d(Y_1, Y_2|X_1) = d(Y_1, Y_2, X_1) - d(X_1) = 1 + 2p - p^2.$$

For the conditional distribution of $P_{y_1,y_2|x_1,x_2}$

$$\begin{aligned}P_{y_1,y_2|x_1,x_2}(y_1, y_2|x_1, x_2) &= (1-p)^2\delta(y_1 - \beta(x_1 + x_2))\delta(y_2 - x_2) + (1-p)p \cdot P_{n_1}(y_1)\delta(y_2 - x_2) + (1-p)p \\ &\quad \cdot P_{n_2}(y_2)\delta(y_1 - \beta(x_1 + x_2)) + p^2 \cdot P_{n_1}(y_1)P_{n_2}(y_2),\end{aligned}$$

the joint distribution $P_{y_1,y_2,x_1,x_2}$ is

$$\begin{aligned}P_{y_1,y_2,x_1,x_2}(y_1, y_2, x_1, x_2) &= (1-p)^2\delta(y_1 - \beta(x_1 + x_2))\delta(y_2 - x_2)P_{x_1}(x_1)P_{x_2}(x_2) + (1-p)p \cdot P_{n_1}(y_1)\delta(y_2 \\ &\quad - x_2)P_{x_1}(x_1)P_{x_2}(x_2) + (1-p)p \cdot P_{n_2}(y_2)\delta(y_1 - \beta(x_1 + x_2))P_{x_1}(x_1)P_{x_2}(x_2) \\ &\quad + p^2 \cdot P_{n_1}(y_1)P_{n_2}(y_2)P_{x_1}(x_1)P_{x_2}(x_2).\end{aligned}$$

The Rényi information dimension is calculated as:

$$d(Y_1, Y_2, X_1, X_2) = 2(1-p)^2 + 3(1-p)p + 3(1-p)p + 4p^2 = 2 + 2p.$$

So the CID is given by,

$$d(Y_1, Y_2|X_1, X_2) = d(Y_1, Y_2, X_1, X_2) - d(X_1, X_2) = 2p.$$

Now we are ready to calculate the MIDs:

$$d(Y_1, Y_2; X_1) = d(Y_1, Y_2) - d(Y_1, Y_2|X_1) = (1-p)^2,$$

$$\begin{aligned}d(Y_1, Y_2, X_1; X_2) &= d(Y_1, Y_2, X_1) - d(Y_1, Y_2, X_1|X_2) = d(Y_1, Y_2, X_1) - (d(Y_1, Y_2|X_1, X_2) + d(X_1|X_2)) \\ &= d(Y_1, Y_2, X_1) - (d(Y_1, Y_2|X_1, X_2) + d(X_1)) = 2 + 2p - p^2 - (2p + 1) = 1 - p^2,\end{aligned}$$

where we have used the fact that $X_1$ and $X_2$ are independent.

## APPENDIX III

### PROOF OF PROPOSITION 3

Proposition 3 can be proved by following the induction in [12] for BEC. In this appendix we only identify the key steps of the proof. Let us define the following terminology. A pair of new real-valued-input-output channels $W':\mathcal{X}\to\mathcal{Y}\times\mathcal{Y}$ and $W'':\mathcal{X}\to\mathcal{Y}\times\mathcal{Y}\times\mathcal{X}$ is said to be obtained from two independent copies of the original channel $W:\mathcal{X}\to\mathcal{Y}$ after a single-step transformation if for all $Y_1, Y_2 \in \mathcal{Y}$ and $X_1, X_2 \in \mathcal{X}$

$$W'(Y_1, Y_2|X_1) = \int W(Y_1|\beta(X_1 + X_2))W(Y_2|X_2)P_{X_2}(X_2)dX_2,$$
$$W''(Y_1, Y_2, X_1|X_2) = W(Y_1|\beta(X_1 + X_2))W(Y_2|X_2)P_{X_1}(X_1).$$

In such case, we use the following notation for $W$, $W'$, and $W''$:

$$(W, W) \mapsto (W', W'').$$

Obviously, the one-step transformation in Fig. 1 can be written in $(W_{2^0}^{(1)}, W_{2^0}^{(1)}) \mapsto (W_{2^1}^{(1)}, W_{2^1}^{(2)})$, where $W_{2^0}^{(1)}$ is the original sparse noisy channel, $W_{2^1}^{(1)}$ is the new channel $W'(Y_1, Y_2|X_1)$, and $W_{2^1}^{(2)}$ is the new channel

$W''(Y_1, Y_2, X_1 | X_2)$. From Proposition 1 and Proposition 2, we know that the MIDs satisfy

$$MID_{2^1}^{(1)} = \left(MID_{2^0}^{(1)}\right)^2,$$
$$MID_{2^1}^{(2)} = 2MID_{2^0}^{(1)} - \left(MID_{2^0}^{(1)}\right)^2.$$

Then for the $(n + 1)$-th step transformation in Fig. 2, we have

$$W_{2^{n+1}}^{(2i-1)}(\mathbf{y}_1, \mathbf{y}_2, X_1^{2i-2} | X_{2i-1}) = \int W_{2^n}^{(i)}\left(\mathbf{y}_1, z_{(1),1}^{i-1} | \beta(X_{2i-1} + X_{2i})\right) W_{2^n}^{(i)}\left(\mathbf{y}_2, z_{(2),1}^{i-1} | X_{2i}\right) P_{X_{2i}}(X_{2i}) dX_{2i}, \quad (31)$$

$$W_{2^{n+1}}^{(2i)}(\mathbf{y}_1, \mathbf{y}_2, X_1^{2i-1} | X_{2i}) = W_{2^n}^{(i)}\left(\mathbf{y}_1, z_{(1),1}^{i-1} | \beta(X_{2i-1} + X_{2i})\right) W_{2^n}^{(i)}\left(\mathbf{y}_2, z_{(2),1}^{i-1} | X_{2i}\right) P_{X_{2i-1}}(X_{2i-1}), \quad (32)$$

where $z_{(1),1}^{i-1} = v_1^{i-1}$ and $z_{(2),1}^{i-1} = v_{1+2^n}^{i-1+2^n}$. Let us note that there is a one-to-one correspondence between $\{z_{(1),1}^{i-1}, z_{(2),1}^{i-1}\}$ and $X_1^{2i-2}$. Therefore, we can write $\left(W_{2^n}^{(i)}, W_{2^n}^{(i)}\right) \mapsto \left(W_{2^{n+1}}^{(2i-1)}, W_{2^{n+1}}^{(2i)}\right)$ after the following substitutions:

$$W \leftarrow W_{2^n}^{(i)}, W' \leftarrow W_{2^{n+1}}^{(2i-1)}, W'' \leftarrow W_{2^{n+1}}^{(2i)},$$
$$X_2 \leftarrow X_{2i}, X_1 \leftarrow X_{2i-1}, Y_1 \leftarrow (\mathbf{y}_1, z_{(1),1}^{i-1}), Y_2 \leftarrow (\mathbf{y}_2, z_{(2),1}^{i-1}).$$

And Proposition 3 follows.

## APPENDIX IV

### PROOF OF THEOREM 2

For Lemma 1 and Lemma 2 below, we consider the binary version of the matrix $\mathbf{G}_0^{\otimes n}$, where the 1's in the real case are replaced by 1's in the binary case. Also, let the rows and columns of $\mathbf{G}_0^{\otimes n}$ be labeled by $0, 1, \ldots, M - 1$, where $M = 2^n$. The following notation is needed to specify the non-zero elements of $\mathbf{G}_0^{\otimes n}$. For any two non-negative integers $i$ and $j$, we write $i \preccurlyeq j$, if for any $k$, the $k$-th digit in the binary representation of $i$ is 1 only if the $k$-th digit of $j$ is also 1.

*Lemma 1.* The $(i, j)$ coordinate of $\mathbf{G}_0^{\otimes n}$ is non-zero if $i \preccurlyeq j$, otherwise it is 0.

Proof. The proof is by induction on $n$. For $n = 1$, the coordinates $(0,0), (0,1)$ and $(1,1)$ are 1 and the coordinate $(1,0)$ is zero. Hence, the lemma holds for $n = 1$. Now, suppose it holds for $n$. For $n + 1$, it is proved by considering four different cases:

Case 1: $i < 2^n$ and $j < 2^n$. In this case $(i, j)$ coordinate of $\mathbf{G}_0^{\otimes(n+1)}$ is equal to the $(i, j)$ coordinate of $\mathbf{G}_0^{\otimes n}$. Therefore, the lemma follows by the induction hypothesis.

Case 2: $i < 2^n$ and $2^n \leq j < 2^{n+1}$. In this case $(i, j)$ coordinate of $\mathbf{G}_0^{\otimes(n+1)}$ is equal to the $(i, j - 2^n)$ coordinate of $\mathbf{G}_0^{\otimes n}$. Notice that in this case, $i \preccurlyeq j$ if and only if $i \preccurlyeq j - 2^n$. Therefore, the lemma follows by the induction hypothesis for $(i, j - 2^n)$.

Case 3: $i \geq 2^n$ and $j < 2^n$. In this case $i \preccurlyeq j$ does not hold and the $(i, j)$ coordinate of $\mathbf{G}_0^{\otimes(n+1)}$ is always 0 due to the recursive structure.

Case 4: $2^n \leq i, j < 2^{n+1}$. In this case, the $(i, j)$ coordinate of $\mathbf{G}_0^{\otimes(n+1)}$ is equal to the $(i - 2^n, j - 2^n)$ coordinate of $\mathbf{G}_0^{\otimes n}$. Also, $i \preccurlyeq j$ is equivalent to $i - 2^n \preccurlyeq j - 2^n$. This completes the proof of the lemma.

In the two following lemmas, we consider the indices $a_0, a_1, \ldots, a_{n-1}$, where $a_j = M - 1 - 2^j$.

*Lemma 2.* Consider the sub-matrix $\mathbf{G}'_{M \times n}$ of $\mathbf{G}_0^{\otimes n}$ which consists of the columns of $\mathbf{G}_0^{\otimes n}$, indexed by $a_j$'s. Then all the rows of $\mathbf{G}'$ are distinct. Namely, they span all the possible binary vectors of length $n$.

Proof. By Lemma 1, the $(i, a_j)$ element of $\mathbf{G}_0^{\otimes n}$ is 1 if and only if $i \preccurlyeq a_j$. Notice that the binary representation of $a_j$ is all 1 except in the $j$-th position (counted from the right, starting from 0). Hence, $i \preccurlyeq a_j$ if and only if the $j$-th digit in the binary representation of $i$ is 0. Therefore, the $i$-th row of the matrix $G'$ is indeed the binary representation of $M - 1 - i$. This implies that all the rows of $G'$ are distinct.

*Lemma 3.* For any $n \geq c$ and $j = 0, 1, \ldots, n-1$, we have

$$MID_{2^n}^{(a_j+1)} \geq 1 - (2p - p^2)^{2^{n-c}},$$

for some constant $c$.

Proof. We choose the integer $c$ such that

$$2p^{2^c} \leq 2p - p^2.$$

The proof is by induction on $n$ for $n \geq c$. For the base of induction $n = c$,

$$MID_{2^n}^{(a_j+1)} \geq MID_2^{(1)} = (1-p)^2 = 1 - (2p - p^2).$$

Now suppose the lemma holds for $n$. For $n+1$, and $j = 1, 2, \ldots, n$, we have $a_j + 1 = 2^{n+1} - 2^j$ and

$$1 - MID_{2^{n+1}}^{(2^{n+1}-2^j)} = \left(1 - MID_{2^n}^{(2^n - 2^{j-1})}\right)^2 \leq (2p - p^2)^{2^{n+1-c}},$$

where the first equality is by the recursion formulas for $MID$ and the inequality is by the induction hypothesis. Also, for $j = 0$,

$$MID_{2^{n+1}}^{2^{n+1}-1} = \left(1 - p^{2^n}\right)^2 \geq 1 - 2p^{2^n} \geq 1 - (2p - p^2)^{2^{n-c}},$$

where the last inequality holds by the particular choice of $c$. This completes the proof of lemma.

*Corollary 1.* Let $\mathbf{A}$ denote the polar linear encoding matrix. Then all the rows of $\mathbf{A}$ are distinct, for $n$ large enough.

Proof. According to the criteria for the selection of good bit-channels and by Lemma 3, all the indices $a_j + 1 = M - 2^j$, for $j = 0, 1, \ldots, n-1$ are among the selected good bit-channel indices, for large enough $n$. By Lemma 2, the intersection of all the $M$ rows with the columns $a_j + 1$ ($a_j$ if indexing starts from 0) of $G^{\otimes n}$ are distinct. This proves the corollary.

Now we are ready to prove Theorem 2. Let us recall that for $SANC(p)$, for channel input $x$, the channel output $y$ is defined as:

$$y = x + \alpha n, \quad \text{where } \alpha = \begin{cases} 0, & w.p. \ (1-p) \\ 1, & w.p. \ p \end{cases},$$

and $n$ is an absolutely continuous random variable independent of $\alpha$.

The polar linear encoding matrix $\mathbf{A}$ is constructed by selecting the columns of the good channels according to MID within the full $M \times M$ polarization matrix denoted by $\mathbf{H}$. According to Proposition 3, $MID$ of $SANC(p)$ obeys the same recursive formula as channel capacity $I$ of $BEC(p)$. As a result, $SANC(p)$ and $BEC(p)$ have the same good channel indices.

Let $L_I$ denote the number of selected columns of $\mathbf{H}$ with higher MID than $1 - \frac{2^{-M^\beta}}{M}$, namely

$$L_I = \left|\left\{i \in \{1, \ldots, M\}: MID_M^{(i)} \geq 1 - \frac{2^{-M^\beta}}{M}\right\}\right|.$$

Notice that this selection of columns is identical to the binary polar code construction for $BEC(p)$, denoted by $C_{p,\beta}$, under the same criteria on the symmetric capacity $I \geq 1 - \frac{2^{-M^\beta}}{M}$. Then by the channel polarization for BEC, the rate of the corresponding polar code $\frac{L_I}{M}$ approaches $1 - p$ while the decoding error is upper bounded by $2^{-M^\beta}$ [28].

After linear encoding the information vector $\mathbf{x} \in \mathbb{R}^{L_I}$ with $\mathbf{A}$, channel output is

$$\mathbf{y} = \mathbf{A}\mathbf{x} + \boldsymbol{\alpha} \odot \mathbf{n},$$

where $\odot$ denotes point-wise multiplication of two vectors. The operation of the analog decoder is as follows. One can write

$$\mathbf{A}_s\mathbf{x} + \boldsymbol{\alpha}_s \odot \mathbf{n}_s = \mathbf{y}_s,$$

where $\boldsymbol{\alpha}_s$ and $\mathbf{n}_s$ are the $L_I$-dimensional subvectors corresponding to an arbitrary row selection $S$ of size $L_I$. We use $T$ to denote the set of row selections $S$ such that the $L_I \times L_I$ matrix $\mathbf{A}_s$ is non-singular.

Consider the set $U$ which is the union of the solutions of $\mathbf{A}_s\mathbf{z} = \mathbf{y}_s$ for all $S \in T$, namely,

$$U = \bigcup_{S \in T}\{\mathbf{z}: \mathbf{A}_s\mathbf{z} = \mathbf{y}_s\}.$$

If $U$ contains a unique element $\hat{\mathbf{x}}$ which is repeated more than once, the decoder outputs $\hat{\mathbf{x}}$. Otherwise, if there is no repetition or multiple repetitions then the decoder declares an error. Now we discuss the decoder's success and error probability.

Notice that if the selection $S$ is such that $\boldsymbol{\alpha}_s = 0$, The equation becomes

$$\mathbf{A}_s\mathbf{z} = \mathbf{y}_s.$$

And since $\mathbf{A}_s$ is non-singular, there is a unique solution for $\mathbf{z}$ which is equal to the message $\mathbf{x}$.

The decoder's failure can be the result of two possible events:

i) There is a solution $\hat{\mathbf{x}} \neq \mathbf{x}$ in $U$ that is repeated at least twice, i.e., there are two selection $S_1$ and $S_2$ with corresponding sub-matrices $\mathbf{A}_1$ and $\mathbf{A}_2$ such that

$$\mathbf{A}_1\mathbf{x} + \mathbf{n}_1 = \mathbf{y}_1 = \mathbf{A}_1\hat{\mathbf{x}},$$

and

$$\mathbf{A}_2\mathbf{x} + \mathbf{n}_2 = \mathbf{y}_2 = \mathbf{A}_2\hat{\mathbf{x}},$$

where $\mathbf{n}_1 = \boldsymbol{\alpha}_{s_1} \odot \mathbf{n}_{s_1}$ and $\mathbf{n}_2 = \boldsymbol{\alpha}_{s_2} \odot \mathbf{n}_{s_2}$ consists of at least one noisy coordinate and some possibly zero coordinates. Since both $\mathbf{A}_1$ and $\mathbf{A}_2$ are non-singular, we can write

$$\mathbf{A}_1^{-1}\mathbf{n}_1 = \mathbf{x} - \hat{\mathbf{x}} = \mathbf{A}_2^{-1}\mathbf{n}_2.$$

Or equivalently

$$\mathbf{A}_2\mathbf{A}_1^{-1}\mathbf{n}_1 = \mathbf{n}_2.$$

Let $S_1'$ and $S_2'$ be the subsets of $S_1$ and $S_2$ which contain the indices corresponding to the non-zero entries of $\boldsymbol{\alpha}_{s_1}$ and $\boldsymbol{\alpha}_{s_2}$, respectively. We know that $S_1'$ and $S_2'$ are non-empty. Consider the following three cases:

Case 1: The subsets $S_1'$ and $S_2'$ are not identical.
Case 2: The subsets $S_1'$ and $S_2'$ are identical and $\mathbf{A}_2\mathbf{A}_1^{-1}$ is not a permutation matrix.
Case 3: The subsets $S_1'$ and $S_2'$ are identical and $\mathbf{A}_2\mathbf{A}_1^{-1}$ is a permutation matrix.

Notice that $\mathbf{A}_2\mathbf{A}_1^{-1}$ is a non-singular matrix which means that it has at least one non-zero entry in each column. Therefore, any non-zero coordinate of $\mathbf{n}_1$ appears with a non-zero coefficient in at least one entry of $\mathbf{A}_2\mathbf{A}_1^{-1}\mathbf{n}_1$. As a result, in Case 1, $\mathbf{A}_2\mathbf{A}_1^{-1}\mathbf{n}_1 = \mathbf{n}_2$ imposes a non-zero linear combination of independent and continuous random variables to be zero. Therefore,

$$\Pr\{\mathbf{A}_2\mathbf{A}_1^{-1}\mathbf{n}_1 = \mathbf{n}_2\} = 0.$$

Similarly, Case 2 also imposes a non-zero linear combination of independent and continuous random variables to be zero which has a zero probability. For Case 3, the equation $\mathbf{A}_2\mathbf{A}_1^{-1}\mathbf{n}_1 = \mathbf{n}_2$ becomes trivial and always holds. However, we established in Corollary 1 that all the rows of $\mathbf{A}$ are distinct. As a result, $\mathbf{A}_1$ and $\mathbf{A}_2$ which are distinct sub-matrices of $\mathbf{A}$ with $L_I$ selection of rows can not be transformed to each other with row permutations. Hence, $\mathbf{A}_2\mathbf{A}_1^{-1}$ cannot be a permutation matrix and Case 3 never happens for large enough $N$ satisfying Corollary 1.

We showed that

$$\Pr\{\hat{\mathbf{x}} \neq \mathbf{x}: \hat{\mathbf{x}} \in U, \hat{\mathbf{x}} \text{ is repeated at least twice}\} = 0.$$

Next, we consider the second event of the decoding failure.

ii)  There is no solution $\hat{\mathbf{x}} = \mathbf{x}$ in $U$ or it is just repeated once.

We relate the probability of decoding failure in this case to the probability of error in decoding the polar code $C_{p,\beta}$ constructed for $BEC(p)$. Notice that any realization of $\boldsymbol{\alpha}$ is equivalent to an erasure pattern for $BEC(p)$ with the same probability. Let $L_c$ denote the number of zero entries of $\boldsymbol{\alpha}$, i.e.,

$$L_c = |\{i: \alpha_i = 0\}|.$$

Also let $\mathbf{A}_s$ denote the $L_c \times L_I$ sub-matrix of $\mathbf{A}$ that contains the rows of $\mathbf{A}$ corresponding to the zero entries of $\boldsymbol{\alpha}$. Notice that the optimum decoding of the erasure pattern $\boldsymbol{\alpha}$ for the corresponding code constructed for $BEC(p)$, i.e., recovering the transmitted message uniquely given the unerased positions, is successful if and only if $\mathbf{A}_s$ is full column rank. Let us note that any $\mathbf{A}_s$ with full column rank on the binary field will also have full column rank on $\mathbb{R}$. This is because $\mathbf{A}_s$ only consists of 0s and 1s. On the binary field, there exists a submatrix of size $L_I \times L_I$ of which the determinant is 1. This implies that the number of 1's in the expansion of $\det(\mathbf{A}_s)$ is odd. Therefore, on $\mathbb{R}$ the number of non-zero elements, which can be $\pm 1$'s, in the expansion of $\det(\mathbf{A}_s)$ is also odd. As a result $\det(\mathbf{A}_s) \neq 0$, implying $\mathbf{A}_s$ is full column rank on $\mathbb{R}$.

Let $P_1$ denote the probability of the failure in decoding the codeword picked from $C_{p,\beta}$ and transmitted over $BEC(p)$. We know that $P_1 \leq 2^{-M^\beta}$ [28]. Let $\mathcal{E}$ denotes the event that $\mathbf{A}_s$ is full column rank and $L_c > L_I$ ($\mathbf{A}_s$ being full-column rank already implies that $L_c \geq L_I$. However we need the inequality to be strict). Then from union bound we have

$$\Pr\{\mathcal{E}^c\} = 1 - \Pr\{\mathcal{E}\} \leq P_1 + P_2, \quad \text{which implies that}$$

$$\Pr\{\mathcal{E}\} \geq 1 - P_1 - P_2, \qquad (33)$$

where $P_2 = \Pr\{L_c = L_I\}$. Let us denote $P_3 = \Pr\{L_c < L_I\}$. We have

$$P_3 > \binom{M}{L_I - 1}(1-p)^{L_I-1}p^{M-L_I+1},$$
$$P_2 = \binom{M}{L_I}(1-p)^{L_I}p^{M-L_I}.$$

As a result,

$$\frac{P_3}{P_2} \geq \frac{L_I}{M - L_I + 1} \cdot \frac{p}{1-p} \to 1, \quad as\ M \to \infty, \tag{34}$$

since $\frac{L_I}{M} \to (1-p)$ from channel polarization for $BEC(p)$ [28]. We note that

$$P_1 \geq \Pr\{L_c < L_I\} = P_3. \tag{35}$$

Combine (33), (34), and (35), we get that

$$\Pr\{\mathcal{E}\} \geq 1 - O\left(2^{-M^\beta}\right).$$

Therefore, the event $\mathcal{E}$ happens with probability at least $1 - O\left(2^{-M^\beta}\right)$. If the event $\mathcal{E}$ happens, we claim that the solution $\hat{\mathbf{x}} = \mathbf{x}$ in $U$ is repeated at least $L_c - L_I + 1 \geq 2$ times. Notice that $\mathbf{A}_s$ is a full column-rank $L_c \times L_I$ matrix and it has at least one $L_I \times L_I$ full-column rank matrix, denoted by $\mathbf{A}_s'$. Furthermore, W.L.O.G one can assume there is no zero row in $\mathbf{A}$. If there is, then one can just simply remove it without any change in the performance of polar code. For each row $\mathbf{c}$ of $\mathbf{A}_s$ not included in $\mathbf{A}_s'$, one can write it as a non-zero linear combination of some of the rows of $\mathbf{A}_s'$. Then by adding $\mathbf{c}$ to $\mathbf{A}_s'$ and removing one of the summands of $\mathbf{c}$ in $\mathbf{A}_s'$, one can get another sub-matrix $L_I \times L_I$ of $\mathbf{A}_s$, which is readily proved to be non-singular. Therefore, there are at least $L_c - L_I + 1 \geq 2$ non-singular sub-matrices of $\mathbf{A}_s$ of size $L_I \times L_I$, each of them yielding a solution $\hat{\mathbf{x}} = \mathbf{x}$ in $U$. This completes the proof of Theorem 2.

## ACKNOWLEDGEMENT
The authors would like to thank the anonymous reviewers for their insights and suggestions.